# Fair by design

A sociotechnical approach to justifying the fairness of AI-enabled systems cross the lifecycle

Marten H. L. Kaas

University of York, marten.kaas@york.ac.uk

Christopher Burr

Alan Turing Institute, cburr@turing.ac.uk

Zoe Porter

University of York, zoe.porter@york.ac.uk

Berk Ozturk

University of York, berk.ozturk@york.ac.uk

Philippa Ryan

University of York, philippa.ryan@york.ac.uk

Michael Katell

Alan Turing Institute, mkatell@turing.ac.uk

Nuala Polo

nualapolo@gmail.com

Kalle Westerling

Alan Turing Institute, kwesterling@turing.ac.uk

Ibrahim Habli

University of York, ibrahim.habli@york.ac.uk

Fairness is one of the most commonly identified ethical principles in existing AI guidelines, and the development of fair AI-enabled systems is required by new and emerging AI regulation. But most approaches to addressing the fairness of AI-enabled systems are limited in scope in two significant ways: their substantive content focuses on statistical measures of fairness, and they do not emphasize the need to identify and address fairness considerations across the whole AI lifecycle. Our contribution is to present an assurance framework and tool that can enable a practical and transparent method for widening the scope of fairness considerations across the AI lifecycle and move the discussion beyond mere statistical notions of fairness to consider a richer analysis in a practical and context-dependent manner. To illustrate this approach, we first describe and then apply the framework of Trustworthy and Ethical Assurance (TEA) to an AI-enabled clinical diagnostic support system (CDSS) whose purpose is to help clinicians predict the risk of developing hypertension in patients with Type 2 diabetes, a context in which several fairness considerations arise (e.g., discrimination against patient subgroups). This is supplemented by an open-source tool and a fairness considerations map to help facilitate reasoning about the fairness of AI-enabled systems in a participatory way. In short, by using a shared framework for identifying, documenting and justifying fairness considerations, and then using this deliberative exercise to structure an assurance case, research on AI fairness becomes reusable and generalizable for others in the ethical AI community and for sharing best practices for achieving fairness and equity in digital health and healthcare in particular.

# 1 INTRODUCTION

To say that work on fairness and AI-enabled systems has exploded would be an understatement. For instance, in 2022 Dolata et al. [12] reviewed 310 papers on algorithmic fairness. An entire subdiscipline of research in the technological ethics space now revolves around issues of fairness and AI-enabled systems, and for good reason. AI-enabled systems are beginning to permeate virtually every sector of the economy [41,49] and are increasingly in our homes and on our bodies [40]. Stakeholders affected by the use of AI-enabled systems should not just be assured of the safety of such systems but also their ethical acceptability. Among those important ethical considerations is the fairness of AI-enabled systems.

Interest in the fairness of AI-enabled systems is so high, in part, because it is an ethical principle that is seemingly amenable to mathematical formalization and is also grounded in legal safeguards (e.g. human rights law, non-discrimination) [21,30,47]. Fairness consistently appears as an ethical principle in existing AI guidelines [15,23] and, in addition to being grounded in already existing legal safeguards (e.g., the UK Equality Act 2010), fairness features prominently in emerging AI-specific regulation, e.g., the EU AI Act, UK AI White Paper and US AI Bill of Rights Blueprint. The fairness of an AI-enabled system (or lack thereof) is fairly easy to assess through statistics-based fair machine learning metrics if, for example, one can demonstrate that some quantifiable measure, such as statistical parity, is satisfied. Moreover, statistical fairness is relatively easy to analyze at an abstract level, away from contextual details and highly specified use-contexts. Even when details are present, statistical fairness is usually only analyzed from the point of view of the prediction-recipient, i.e., from the point of view of the person receiving a prediction generated by an AI-enabled system. In [47] for example the focus is on whether a prediction-recipient, a married woman named Alice, is treated fairly by an AI-enabled classifier that labels Alice as having good or bad credit and hence whether she is deserving of a loan or not. Unsurprisingly, it is not clear whether the AI-enabled classifier treats Alice fairly, or is generally fair to the prediction-recipient given their gender, because different statistical notions of fairness are satisfied (e.g., conditional statistical parity) while others are not satisfied (e.g., statistical parity) [47].

This brings us to the first of two limitations of most current work on AI fairness, which have already been acknowledged by others in the literature [5,43,48]. Current approaches are narrow in scope in two different ways. The first is the narrowness of their substantive content. Fairness, a rich philosophical concept intimately linked to justice, mutual respect and what we owe each other, and important for the well-being and flourishing of people in society [37,42], is frequently reduced to some statistical measure that can be satisfied via different technical solutions. In short, normative questions such as, "What notion of fairness ought we strive to satisfy?" or "How should affected stakeholder groups be involved in mitigating bias?" are often left unaddressed or overlooked (often for understandable reasons, such as resource limitations or lack of skills and capabilities within project teams).

Enter the second limitation. Even if specific notions of fairness are recommended, essentially nothing is said about when in the AI lifecycle issues of fairness ought to be addressed. Most current work on AI fairness gives too little consideration to temporality, i.e., when or at what point (or points) in the AI lifecycle issues of fairness ought to be considered.[1] The prevailing underlying assumption appears to be that fairness ought to be assessed either pre-deployment or immediately post deployment. While this approach seems intuitive, i.e., there are certain natural pause points or milestones in a project's lifecycle where one might assess the system's fairness, these assessments are limited in their usefulness.

Our goal in this paper is to introduce trustworthy and ethical assurance – a framework and open-source tool – and demonstrate how trustworthy and ethical assurance can enable a practical and transparent method for widening the scope of fairness considerations across the AI lifecycle. Trustworthy and ethical assurance is designed to support the development of a structured argument, known as an assurance case, which provides reviewable (and contestable) assurance that a set of claims about a normative goal (e.g., fairness) of a data-driven technology are warranted given the available evidence [7,36]. This approach is advantageous because it can help both to systematize the identification, documentation and justification of fairness considerations as well as support efforts for reproducibility. Importantly, our framework enables reasoning about more substantive sociotechnical fairness considerations by widening the scope of the discussion beyond mere technical considerations. To illustrate this, we use our framework to give a richer analysis of fairness in a context-dependent manner, specifically by looking at the case of an AI-enabled CDSS (clinical diagnostic support system). Additionally, our framework allows for the identification and documentation of fairness considerations throughout all the different stages of the AI lifecycle and not just at certain milestones. We therefore also include a fairness considerations map to help affected stakeholders navigate salient fairness considerations in the healthcare context and to facilitate the process of assuring fairness across the AI lifecycle using our framework and open-source tool.[2]

This paper is organized as follows. In section 2 we briefly summarize the literature on fairness and AI before describing the limitations in the current literature in more depth.[3] In section 3 we introduce the idea of trustworthy and ethical assurance, our case study of an AI-enabled CDSS and an AI lifecycle model that can facilitate reasoning about fairness considerations across all stages of an AI-enabled system's lifecycle. In section 4 we draw together the pieces introduced in section 3 to demonstrate how one might assure fairness of an AI-enabled CDSS across the lifecycle in a systematic and reproducible way that goes beyond the technical to include socio-political considerations of fairness.

## 2 FAIRNESS AND AI

As mentioned above, research in the fairness of AI-enabled systems is growing exponentially. It is worth drawing out a few pervasive trends in the literature to help frame the subsequent sections and the value trustworthy and ethical assurance has in assuring fairness of AI-enabled systems across the lifecycle.[4] Importantly, it must be noted that we are not the first to acknowledge these trends nor the first to address them (e.g., by using model cards) [32]. Like many before us, our hope is that this work encourages others to participate in thorough fairness assurance across the AI lifecycle.

### 2.1 Formalizing Fairness

One pervasive trend is to define and measure fairness in mathematical terms, and this has obvious benefits. First, both unsophisticated and more sophisticated notions of fairness are readily amenable to mathematical formalization. For example, one might think that fairness merely amounts to binary predictions (generated by an AI-enabled system) that are statistically independent of a given sensitive attribute, e.g., gender. In other words, fairness is said to obtain when the probability of a certain outcome (e.g., approval for a loan, risk of recidivism, etc.) is the same across different groups, where groups may be based on protected or sensitive characteristics like gender mentioned above. This is known as statistical parity and its advantages and disadvantages have been discussed extensively throughout the literature [21,30,35].

---

[1] A notable, recent exception is the policy guidance put out by Australia's National AI Centre (NAIC), which directly explores how practices designed to operationalize their 8 principles for responsible AI can be situated at different stages of a project's lifecycle [38].
[2] However, we should note that the tool and framework are not restricted to use within the health or healthcare domain specifically. More information about them can be found here: https://github.com/alan-turing-institute/AssurancePlatform.
[3] Those familiar with the literature could safely skip this section.
[4] For recent reviews of fairness notions in AI and machine learning see [30,35].

A more sophisticated notion of fairness might entail analyzing more of the details embedded in the data, for example the false positive rates and/or the false negative rates. One might reason, for example, that fairness amounts to equal false positive rates when comparing different groups. That is, fairness is said to obtain when the probability of an actual "negative" outcome (e.g., being rejected for a loan) being incorrectly classified as a "positive" outcome (e.g., being approved for a loan) is the same across different groups, where again these groups may be based on protected or sensitive characteristics. This is known as predictive equality[5] and, as with statistical parity, its advantages and disadvantages have been extensively discussed throughout the literature [21,30,35].

A second benefit is that the mathematical formalization of fairness can allow one to address issues of fairness through technical solutions. For instance, an AI-enabled system used for criminal risk assessment that makes a binary recidivism prediction (e.g., whether in the next year a person will reoffend or not) can be made "fairer" if the system is retrained on different data or data processed in a different manner. Another approach might involve hard-coding certain restrictions to ensure fair outcomes at the system level despite a biased model. So, for example, if there is a historical bias in the hiring trends in the information technology sector and a predictive model trained on this data discriminates against female applicants, developers could hard code something along the following lines.[6] For every male applicant identified as a potential new hire, a female applicant must also be selected as a potential hire. This however reveals the insidiousness of applying only technical solutions to address issues of fairness. It can be quite easy to game formalized notions of fairness, especially statistical parity, by artificially selecting for or against individuals from certain groups of people to meet fairness requirements, at least on paper [3,30].

To be clear, this reduction of fairness to some statistical measure is understandable. Statistical measures of fairness enable researchers and developers to quantify the impact or effect of specific actions undertaken during the course of a system's development. And, such quantification can be a powerful means of justification or assurance. However, it is well known how statistics can belie the complex set of actions or phenomena that give rise to the formal measure in the first place. As mentioned above, artificially selecting for individuals from certain groups can result in pernicious self-fulfilling prophecies [13] and the amplification of systemic discrimination against already marginalization communities [17]. This can be succinctly stated as follows.

> "[S]tatistical facts are often facts about how we choose to act. Since we can morally evaluate how we choose to act, we cannot simply take statistical facts for granted when justifying policies: we need to ask the prior question of whether it can be justified that we make these statistical facts obtain." [29]

So, while it is understandable to reduce fairness to some statistical measure, the more important question is whether this is a defensible approach, especially if fairness is only analyzed according to statistical measures. Our view, which we expand on in sections 3 and 4, is that a reductionist approach is simply insufficient to capture all the relevant fairness considerations.

---

[5] It is also known simply as equal false-positive rates [21].
[6] Amazon's approach was to simply scrap their AI-enabled hiring system altogether.

## 2.2 Focusing On Prediction-Recipients

Another pervasive trend in the literature is to focus primarily, if not exclusively, on the fairness of an AI-enabled system insofar as it affects one stakeholder group, typically the prediction-recipient. And again, there are numerous benefits to this kind of approach. Focusing only on the prediction-recipient greatly simplifies considerations of fairness. So long as unwanted bias against prediction-recipients has been dealt with, why complicate the issue by considering the disparate impacts an AI-enabled system may have on other affected stakeholders? Not only does focusing on a single affected stakeholder group simplify things for the developer, it also simplifies things for other involved stakeholder groups, e.g., regulators, service providers, and lawyers/the courts.[7] This is perhaps most pronounced in the healthcare sector wherein effects on the patient, i.e., prediction-recipient, are of paramount importance. Even if the clinician, i.e., the expert user, is brought into the discussion, it is usually in the context of how interactions between clinicians and an AI-enabled system can result in disparate impacts to certain patient groups [18]. Fairness considerations concerning the clinicians themselves are often left by the wayside [25].

Another benefit that arises from the focus on a single affected stakeholder group, again typically the prediction-recipients, is that assessments of fairness can be temporally localized to specific milestones in the AI lifecycle. In particular, fairness assessments can be temporally localized to the point between development and deployment. This is a natural milestone in a project lifecycle and provides developers with the opportunity to pilot the system and make changes to improve both its performance and usability. It is common, but highly inappropriate, to reason in the following way. Since prediction-recipients will only interact with the system once it has been deployed, assessments of the AI-enabled system's fairness can be reasonably ignored until the point in the lifecycle when that interaction becomes imminent. So, fairness of the AI-enabled system need only be assessed once the system has been designed and developed. Of course, such reasoning is inappropriate because it invariably reduces fairness to some statistical or technical measure that can be addressed only through statistical or technical solutions. Instead of addressing potentially discriminatory effects on prediction-recipients at an earlier design stage, developers often resort to treating the symptoms rather than the root causes of unfair outcomes.

As with the formalization and reduction of fairness described above, focusing on the fairness of an AI-enabled system insofar as it affects one stakeholder group may be understandable, but it is an open question as to whether it is defensible. The evidence appears to be suggesting that, in complicated environments wherein multiple different stakeholder groups interact with an AI-enabled system in different ways (e.g., in healthcare), a focus on only one affected stakeholder group is insufficient to ground claims that a given AI-enabled system is fair [25].

## 2.3 Lack of Systematic and Reproducible Processes

A third trend is the lack of systematic and reproducible processes available for explaining the actions, arguments and evidence which justify confidence that systems are acceptably fair in their operating contexts. This trend is largely a consequence of the two trends just highlighted, namely the formalization of fairness and a focus on prediction-recipients. While there is some guidance in the literature as to which statistical measure of fairness is appropriate given the intended use of an AI-enabled system and other contextual factors, there is hardly a widely agreed upon systematic approach to choosing a particular formalization of fairness let alone a systematic approach to identifying and documenting relevant fairness considerations [30]. When coupled with a focus on fairness as it concerns prediction-recipients, the process of creating a "fair AI-enabled system" becomes a veritable Choose Your Own Adventure exercise for developers or procurers of an AI-enabled system. Nowhere is this clearer than in the case of the AI-enabled COMPAS (Correctional Offender

---

[7] We find it useful to distinguish between at least eight different stakeholders groups identified by [31]. These include prediction-recipients, end users, expert users, regulatory agencies, service providers, accident and incident investigators, lawyers and the courts and insurers.

Management Profiling for Alternative Sanctions) system used to predict recidivism. A ProPublica report [1] influentially argued that the system was discriminatory because it yielded unequal false-positive and false-negative rates for black defendants and white defendants (i.e., black defendants were more likely to be incorrectly labelled as high risk while white defendants were more likely to be incorrectly labelled as low risk) [21]. Almost immediately however a rejoinder was published [16] in which it was argued that COMPAS is not racially biased because a given COMPAS score approximately corresponds to the same likelihood of recidivism regardless of race, i.e., COMPAS is fair because it satisfies equal predictive accuracy and calibration [21,30]. Naturally, the company that developed COMPAS argued that its predictions are not racially biased [10], but their system remains inaccessible to the public as does their reasoning and justification for choosing to satisfy certain measures of fairness for prediction-recipients rather than any of the other measures.

## 3 ASSURING FAIRNESS ACROSS THE AI LIFECYCLE

Purely statistical approaches to fairness, while attractive, are nevertheless deficient. There is, however, growing recognition that both sociotechnical approaches to fairness [12] and through-life or end-to-end approaches to fairness [28] are required to address the limitations of most current approaches outlined in the previous section. In particular, there is a need to both widen the scope of fairness considerations in terms of their substantive content and widen the scope of fairness considerations in terms of when or at what point in the AI lifecycle assessments of fairness are made and addressed.

Moreover, there is a need to systematize the identification and documentation of fairness considerations for AI-enabled systems. Not only would such systematization contribute to producing auditable artefacts to ensure compliance with fairness requirements, but it would also support efforts for reproducibility. That is, by using a shared model and framework for identifying and documenting fairness considerations, and then using this deliberative exercise to structure an assurance case, the work becomes reusable and generalizable for others in the community (e.g., sharing best practices for fairness and equity in digital health and healthcare).

Let us unpack this approach by first examining assurance cases.

### 3.1 Trustworthy and Ethical Assurance

What is an assurance case and, moreover, what is trustworthy and ethical assurance?

'Assurance' generally refers to the activity of providing a warrant or establishing justified confidence for accepting a certain claim – such as that the design, development and deployment of a system meets a desired goal like fairness – on the basis of argument and evidence [20].[8] Importantly, we can distinguish between assurance in general, i.e., the activity just described, and the assurance case methodology, i.e., argument-based assurance, which is one of several methods one could use when engaging in the activity of assurance. When developing safety-critical systems, for example, developers have to assure the safety of their system – a task that is often accomplished through the use of a safety case, i.e., appropriately structured arguments justified by defeasible evidence [26,44]. More recently, there has been an interest in applying the same process to assure other claims, that an AI-enabled system is not merely safe, but also *ethically acceptable* [7,36]. This is evidenced, in part, by the work being carried out in the UK regulatory space by the Centre for Data Ethics and Innovation (CDEI).[9] Though it must be noted, the CDEI is interested in assurance in the sense of building trust or confidence in AI-enabled systems and not necessarily in the sense of an assurance case such as is described below.

---

[8] Note that the claims that one might try to assure often revolve around a property (or properties) of a system, e.g., its safety, robustness, etc.
[9] See the CDEI's roadmap to an effective AI assurance ecosystem: https://www.gov.uk/government/publications/the-roadmap-to-an-effective-ai-assurance-ecosystem

An *assurance case* is an instantiated and structured argument and accompanying evidence intended to support a particular claim about a particular system (e.g., that an AI-enabled CDSS is safe). A trustworthy and ethical assurance case is, therefore, a type of assurance case that instantiates a structured argument alongside the evidence intended to support the claim that a particular system is trustworthy and ethically acceptable. In short, trustworthy and ethical assurance is a structed approach to the communication of reasons and evidence about a system that helps affected stakeholders understand and evaluate the claims made about some normative property or goal of the system. To help facilitate the creation of trustworthy and ethical assurance cases, we have developed both an open-source tool and a framework.

First, Trustworthy and Ethical Assurance (TEA) is an open-source tool that can be used by affected stakeholders, though it may primarily be of interest to developers, to help support the process of developing an assurance case to substantiate claims about a given system. But second, this tool is an extension of a comprehensive framework comprised of a set of resources and an argument-based methodology. These resources and methodology play a crucial role in providing a systematic and reproducible framework. Indeed, it is the TEA methodology that contributes most to the systematization and reproducibility of evaluations of an AI-enabled system's fairness, so we focus on that now.

As alluded to above, TEA is an example of what is known as argument-based assurance because it is through the use of an appropriately structured *argument* that a claim, e.g., that an AI-enabled CDSS is fair, is *assured*. While this is certainly not the only way to provide assurance or justify trust,[10] TEA emphasizes the importance of an accessible and structured argument. This is because TEA has its roots in the tradition of informal argumentation that attempts to capture certain shared attitudes about argumentation. These are chiefly a focus on the natural language arguments that are used in public discourse, on understanding argumentation as a dialectical process and on articulating standards, norms or advice for argument evaluation that are not merely captured by the categories of deductive validity, soundness and inductive strength [24].

Though there are different approaches to argument-based assurance [19], there are three basic elements of a TEA case, which are illustrated in Figure 1. First is the top-level claim which, in our case, is the claim that an AI-enabled CDSS is fair. This top-level claim is supported by additional intermediate claims. These intermediate claims are the second element, which collectively specify and operationalize the top-level claim. There can be multiple levels of these intermediate claims, each of which is more specific than the claims in the level above. An example of an intermediate claim that might directly support the top-level claim is, "The AI-enabled CDSS does not discriminate against patients." Note however that this intermediate claim is still quite general and ambiguous. It therefore ought to be further decomposed into additional intermediate claims such as, "The AI-enabled CDSS does not discriminate against marginalized patient sub-groups."

---

[10] The UK's Centre for Data Ethics and Innovation (CDEI) for example has developed an AI assurance guide to help organizations understand how different assurance techniques can be applied to AI-enabled systems.

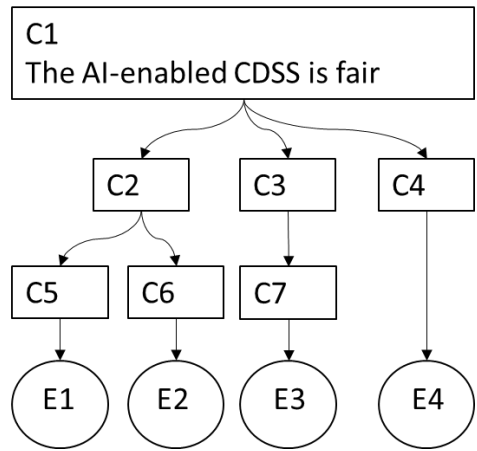

Figure 1: A simple illustration of the three elements in a TEA case. C1 is the top-level claim which is decomposed into the intermediate claims C2-C7. Note that there are multiple levels of intermediate claims, e.g., C2 decomposes into the intermediate claims C5 and C6. Grounding the argument are the pieces of evidence E1-E4, each of which directly supports the intermediate claims it is connected to.

This hierarchical decomposition of claims should proceed in a systematic way until, at the bottom, grounding the entire argument, is the evidence. The evidence is the third element of a TEA case. This evidence should clearly justify the claim(s) it is directly supporting. If, for example, one requires specialist knowledge to understand the connection (or the inference) between the evidence and the claim, then that connection should be included in the TEA case and documented in such a way that a lay audience could understand. This evidence can take many different forms and should do so depending on the claim(s) that evidence is intended to support. For example, relevant documents and transcripts/logs of meetings may suffice as evidence to justify claims that an appropriately diverse team was involved in the design of the AI-enabled CDSS whereas calculated measures of accuracy may suffice as evidence to justify claims that identified statistical measure of fairness were satisfied. Additionally, this style of structured reasoning may enable certain stakeholders to challenge the strength or reliability of these items of evidence, a mindset that is encouraged for the assurance of safety-critical systems in clinical applications [45].

In summation, through TEA we can meaningfully assure the fairness of AI-enabled systems in a systematic and reproducible way. Systematic because the top-level claim about an AI-enabled system's fairness must be hierarchically decomposed, justified by evidence, and ultimately defended as a reasonably argued claim. And reproducible because TEA cases can support scaffolding communities of practice, i.e., facilitate open and transparent reasoning about AI-enabled systems through assurance cases as a way to develop clear standards for those committed to developing fair AI-enabled systems in particular, and trustworthy and ethical AI-enabled systems in general.

### 3.2 The AI Lifecycle

There are many ways in which one might carve up the AI lifecycle space. In general, the lifecycle of a product or system is an abstract overview of that system from its beginning to its end. In practice, however, there are at least two separate considerations that complicate the process of developing a useful lifecycle model for AI-enabled system.

The first consideration is that of granularity. At what level of detail ought one construct the AI lifecycle? At one end of the spectrum are AI lifecycle models that take a fine-grained approach by considering only the machine learning aspects of an AI-enabled system. [2], for example, differentiating between four different stages of the machine learning lifecycle: data management, model learning, model verification and model deployment.

At the other end of the spectrum are AI lifecycle models that take a coarse-grained approach by considering, well, almost everything. [28], for example, also differentiates between four different stages in the AI lifecycle: the world, data, design and the ecosystem. In the first stage, the world, there is a need to understand that AI-enabled systems are produced in an already existent world, one that has certain social and historic realities with preexisting patterns of discrimination and social injustice [28]. The understanding generated in the first stage should inform decisions made in the second stage, data, in which the focus is on the information that is collected [28]. In particular, close attention ought to be paid to the ways in which collected information can encode patterns of discrimination. The information itself, the labels used to organize the data, features of the data that are deemed useful and cognitive biases that influence our interpretation of the data are all important to consider at this stage, among other relevant aspects. This is because the data will ultimately have consequences for the third stage, design, in which the focus is on the development of the AI-enabled system [28]. This stage includes making choices about the learning algorithm, hyperparameters, training and testing datasets, verification and validation benchmarks as well as how the system will be integrated and embedded in other systems. Once deployment occurs we reach the fourth stage, the ecosystem, wherein the focus is on how AI-enabled systems influence and are utilized within the wider social, political, cultural and legal milieu [28]. As many have already warned, analysis of AI-enabled systems in these wider contexts can reveal that they further entrench or exacerbate existing inequities. Moreover, new data generated in part by AI-enabled systems can generate positive feedback loops that amplify existing patterns of discrimination such as in cases of predictive policing [17].

This brings us to the second consideration that complicates the development of a useful AI lifecycle model, namely that the development of AI-enabled systems is not a linear process. AI-enabled systems are not simply developed, their safety (or other property) certified, and then deployed without monitoring or assessment. Rather, AI-enabled systems are developed through an iterative and cyclic development process. As alluded to above, the outputs of AI-enabled systems become part of the world and can therefore be swept up in data collection that will be used to design even more AI-enabled systems, hence their cyclic development. At the same time AI-enabled systems are developed iteratively, and this is especially clear at a finer level of granularity. Developers might, for example, reach the model verification stage before iteratively improving the system by revisiting the model learning or data management stages. All of this complicates the development of an AI lifecycle model. But these considerations aside, an AI lifecycle model can play a critical role in helping to identify at what stages it is appropriate to make and justify certain claims about the system, including claims about fairness, e.g., that an AI-enabled system was designed by an appropriately diverse team or satisfied an identified statistical measure of fairness as mentioned in section 3.1.

Our view is that, when it comes to trustworthy and ethical assurance, an intermediate level of abstraction is required. In particular, the AI lifecycle advanced by [7] is at a sufficient level of abstraction to facilitate effective reasoning about fairness of an AI-enabled system. This lifecycle model, adopted below in Figure 2, is divided into three overarching phases each of which are further divided into four stages [6].

The first phase, project design, is divided into the following four stages: project planning; problem formulation; data extraction or procurement; and data analysis. The goals in the project planning and problem formulation stages revolve around preliminary scoping activities that help to determine the aims and objectives of the project (of developing an AI-enabled system) and to explicate the precise problem that the AI-enabled system is intended to address. In the data extraction or procurement and data analysis stages, the focus shifts towards preliminary assessments of data quality and its provenance as well as its suitability for the problem just identified.

The second phase, model development, is divided into the following four stages: preprocessing and feature engineering; model selection and training; model testing and validation; and then model documentation. After preliminary data analysis, choices must be made about how to modify the data, which occurs in the preprocessing and feature engineering stage, and what kind of model or training algorithm/regime should be used to facilitate effective training of the system, which occurs in the model selection and training stage. Once the model is trained, the next stage involves testing and validating the model. Finally, the goal in the model documentation stage is to ensure that development of the model has been properly documented. This includes documentation that summarizes the model's performance, the evaluation metrics used, as well as information about the model itself (e.g., what learning algorithms were used, or whether an already existing model was adapted for use in the project).

The third phase of the AI lifecycle, system deployment, is divided into the following four stages: system implementation; user training; system use and monitoring; and model updating or deprovisioning. After development, in the system implementation stage, the learned model is produced and integrated into the larger system and its operating environment in which it is just one component. Once implemented, use of the AI-enabled system may require user training and will, moreover, require a certain level of monitoring post-deployment to ensure that it is operating as intended in the environment. Ideally, monitoring would be continuous to inform decisions about whether the system requires updating or deprovisioning.

Before moving on, it is worth mentioning that this lifecycle is both cyclic and iterative. Significant updates to an AI-enabled system may prompt affected stakeholders to revisit the initial project planning and problem formulation stages to reconsider the design of the AI-enabled system and whether it is an appropriate intervention. Within the lifecycle, feedback during the user training stage may prompt developers to iteratively improve the system by returning to the system implementation stage or an even earlier stage in the lifecycle. Before describing how this, i.e., assuring fairness across the AI lifecycle, might happen in practice we turn now to describe our case study, an AI-enabled CDSS.

**3.3 A Clinical Diagnostic Support System**

The use of AI-enabled systems in healthcare, as in all sectors, is increasing and there are continuous efforts to expand their capabilities so that they are able to assist or replace humans. Our case study concerns the use of an AI-enabled CDSS (clinical diagnostic support system) that is intended to assist clinicians in their assessment of patients diagnosed with Type 2 diabetes [33].

Briefly, Type 2 diabetes occurs when a person's pancreatic cells are unable to produce enough insulin, the hormone that regulates blood sugar, or when a person's body cells do not react to insulin, leading to an imbalance in blood sugar levels. Type 2 diabetes can occur at any age and the global prevalence of this health condition has increased gradually [33]. It is predicted that over half a billion people will be diagnosed with Type 2 diabetes by 2040 [11,33]. While Type 2 diabetes can be managed, poor management of the condition can cause the disease to progress and comorbidities, i.e., additional illnesses, to develop, such as blindness, kidney failure, hypertension and heart attacks, strokes, and the need for limb amputation, among others [33].

The early diagnosis of Type 2 diabetes is therefore important not only for the management of the disease itself, but also for the management of the other illnesses that diabetes can cause. This is where AI-enabled systems enter the picture. The basic idea is that AI-enabled systems can be used to advise clinicians in supporting the management of a patient's Type 2 diabetes by providing clinicians with a prediction about whether a patient is at risk of developing certain comorbidities. In particular, our focus is on an AI-enabled system that will assist clinicians by making predictions about whether a patient is at high or low risk of hypertension (e.g., within the next six months when they have another checkup) [39]. Importantly, these are patients that have already been diagnosed with Type 2 diabetes.

Using the terminology introduced above in section 3.2 on the AI lifecycle, we begin with the first stage of the model development phase, preprocessing and feature engineering. To train the AI-enabled system, electronic patient data was collected and preprocessed such that, out of 476,000 patient records, 42,000 were selected for training and testing purposes [33]. From these 42,000 patient records, the 20 most frequent variables (e.g., body mass index measurements, total white blood cell count, red blood cell count, etc.) were included as features for the training and testing data sets [33]. That is, out of all the different pieces of information collected during a patient's visit with their clinician, e.g., blood pressure, heart rate, urea levels, etc., only 20 of the most frequently collected pieces of information were used for training and testing. For the actual training, four different machine learning algorithms were employed: a naïve bayes, a neural network, a random forest and a support vector machine [33]. While each algorithm was used to train a separate model, the four models were combined using a generalized linear model as an ensemble method to combine the predictions of each of the four individual models and thereby increase the performance of the system [33]. The performance metrics Accuracy and Cohen's Kappa values were used as they are commonly utilized to evaluate the performance of ML-based classification models [33]. Accuracy is a metric that measures the correctness of the model, and Cohen's Kappa is a statistical measure that shows the level of agreement between the actual and predicted values by the ML model(s) [33]. Observing higher Accuracy and Cohen's Kappa values corresponds to better performance outputs [33]. In addition, to make the results more explainable, feature importance levels of each input feature have been calculated. The feature importance levels show the relative contribution of each input feature for predicting the output [33].

## 4 ASSURING AI FAIRNESS IN HEALTHCARE

In the previous section we introduced the TEA framework, the AI lifecycle model and a case study of an AI-enabled CDSS. Our aim in this final section is to connect these pieces to demonstrate how one might assure through-life fairness for an AI-enabled CDSS in a systematic and reproducible way, which includes richer socio-political considerations of fairness rather than merely technical considerations of fairness.

### 4.1 The Healthcare Context

Ethics is deeply rooted in healthcare and the practice of medicine. From the Hippocratic oath in antiquity to the principles of modern biomedical ethics as advanced by Beauchamp and Childress [4], it is clear that *doing the right thing* is paramount in the healthcare context. Here, 'doing the right thing' includes considerations of fairness. But what counts as fair can vary both within healthcare contexts, as well as between healthcare contexts and others (e.g., education).

At the heart of the healthcare context is a relationship between two people, a clinician (i.e., nurse, physician, counsellor, medical assistant, etc.) and a patient. Decision making in this relationship is complex even before the introduction of an AI-enabled CDSS. When deciding whether a patient with Type 2 diabetes is in need of hypertension treatment, a clinician needs to be able to assimilate information from multiple sources including electronic and non-electronic patient data, information obtained in conversation with the patient and information gleaned from the patient context and their behaviour [39]. With the addition of an AI-enabled CDSS in the clinical workflow, the clinician also has to consider the CDSS's prediction (e.g., that the patient is at high risk of developing hypertension in the next six months) as well as the CDSS's generated explanation of why it made that prediction [39].[11] The research question is how can we assure fairness, in a meaningful way, throughout the design, development and deployment of the AI-enabled CDSS for all affected stakeholders?

---

[11] The inclusion of an explanation is generally thought to increase trust in the system and facilitate uptake of such systems.

### 4.2 Through-Life Fairness Assurance: Mapping Out Fairness Considerations

As shown in section 3.1, there are three basic elements of an assurance case: (1) a top-level claim, e.g., that an AI-enabled CDSS used to predict whether patients with Type 2 diabetes are at risk of developing hypertension within the next six months is fair; (2) a set of supporting claims that together specify and operationalize the top-level claim, e.g., that appropriate measures of statistical fairness were chosen for testing and validation of the system and that the system provides human-centric explanation of its predictions; and (3), at bottom, the evidence that grounds the argument by justifying the intermediate supporting claims, e.g., documentation that indicates that the CDSS did not violate identified statistical measures of fairness during the testing and validation stage of the lifecycle. By utilizing the TEA methodology we can assure the fairness of an AI-enabled CDSS across its lifecycle and, moreover, do so in a systematic and reproducible way. To help with this process, we have developed a fairness considerations map (see Figure 2) that can facilitate reasoning about the fairness of an AI-enabled system at a high level.

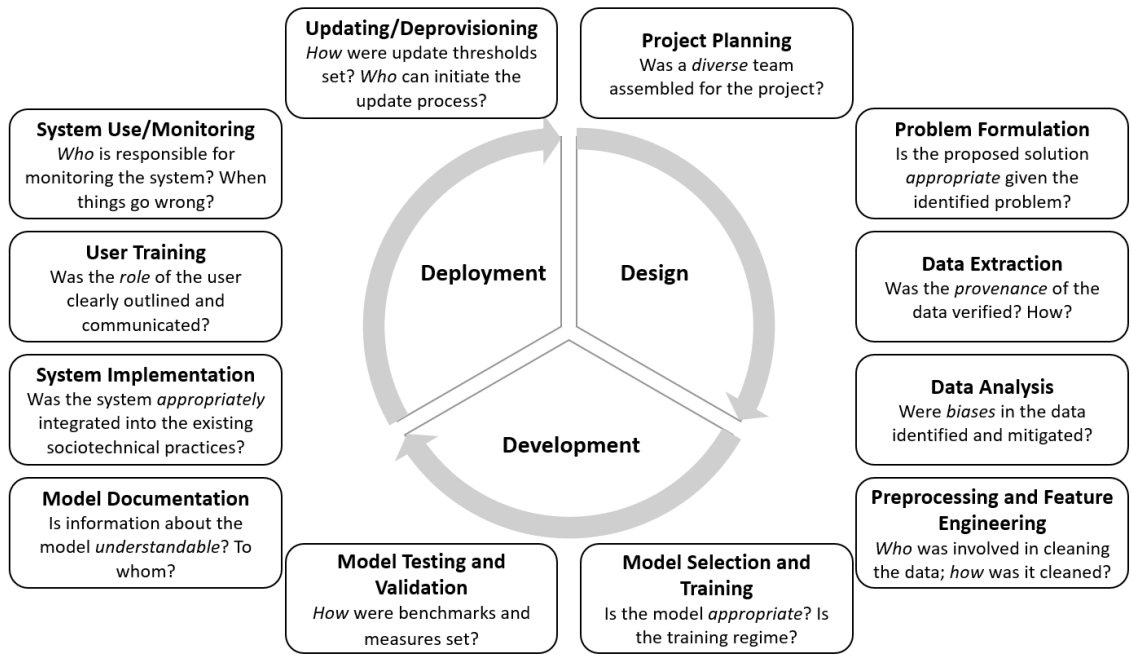

Figure 2: Important intermediate claims that should be taken into consideration when developing a TEA argument to justify the fairness of an AI-enabled system.

Importantly, this map is not exhaustive but rather a proof of concept and guide to prompt further reflection and development of the TEA case. Using the fairness considerations highlighted in Figure 2, we turn now to demonstrate how one might use TEA to assure the fairness of the AI-enabled CDSS we have introduced, beginning with the design phase.

*4.2.1 Fairness Assurance: The Design Phase*

Assuring the fairness of an AI-enabled CDSS ought to begin long before data is collected, analyzed and an ML model trained. It should begin in the design phase, receiving proper attention in the project planning and problem formulation stages. Both stages require *appropriately diverse* teams, and evidence thereof, to justify the top-level goal that the AI-enabled CDSS is fair. Not only is a diverse project team, or inclusive forms of stakeholder engagement and meaningful participatory design, an important precondition for assuring the fairness of an AI-enabled system, it is often required in the healthcare sector in which patient well-being is held as paramount.

Beyond patient well-being, appropriately diverse project teams may also identify that an AI-enabled CDSS may have consequences for clinician well-being. The proposed solution, introducing an AI-enabled CDSS to supply an additional recommendation and thereby increase clinician accuracy when diagnosing hypertension in Type 2 diabetes patients, may actually introduce auxiliary problems (e.g., a distrust in clinicians or a deskilling of clinicians overly reliant on the system). Relatedly, the proposed solution may also exacerbate and further entrench harmful practices (e.g., the reduction of complex biological facts to a single measure of BMI). Furthermore, though the elimination of injustice is not guaranteed, appropriately diverse project teams can help to ensure that marginalized voices are heard and thereby alleviate some of the harms that arise as a result of epistemic injustices such as undervaluing, overlooking and misrepresenting people's testimony [8,14,34]. Prediction-recipients (i.e., the patients) and expert users (i.e., the clinicians) also often speak a different language than the developers. By working together with developers in the earliest stages of the project, patients and clinicians can help mitigate the perpetuation of distorted narratives about their lived realities that are often embedded in technological solutions.

It is equally important in the design phase to ensure that the data is of sufficiently high quality and that biases have been identified and mitigated in the data extraction/procurement and data analysis stages respectively. In our case, the data used was from the Connected Bradford Primary Care dataset [33]. While the dataset contains over one million patient entries, initial filtering was for those patients with Type 2 diabetes resulting in the 476,000 patient records mentioned above in section 3.3 which was ultimately whittled down to a subset of 42,000 patient records [33]. Additionally, because not all patient records were complete, i.e., containing all of the same recorded variables from clinical visits, synthetic data was generated to fill these missing values before training of the ML model [33]. Evidence that fairness has been addressed at these stages in the lifecycle is crucial because biases in the data, e.g., over representation of a particular patient group or data of poor quality, and biases from missing elements in the data, e.g., missing age groups, can have disproportionately harmful effects on different groups of patients. Failure to substantiate claims that fairness concerns have been addressed in these latter stages in the design phase can have a cascading and pernicious effect on the fairness of the system as it is further developed in the development phase [39].

*4.2.2 Fairness Assurance: The Development Phase*

As indicated in Figure 2, there are important elements of fairness in the development phase that are beyond the merely technical or statistical. For example, justifying the fairness of an AI-enabled CDSS in the preprocessing and feature engineering stage ought to involve demonstrating that the data was appropriately cleaned by an appropriate team. In our case, a clinician was consulted at multiple points in the preprocessing and feature engineering stage to ensure that the chosen features from the patient records align with those features a clinician uses to assess a Type 2 diabetes patient's risk of developing hypertension. Additionally, missing data was imputed as the average from other patient records in which the value was present [33].[12] Justification concerning the sufficiency of this particular process would need to be included in the TEA case since there are myriad ways in which developers can choose to address the problem of missing data, each with different consequences on the fairness of the model and overall AI-enabled CDSS.

---

[12] So if, for example, the BMI was missing in a patient record, it would be filled in with the average calculated from BMI's recorded for other patients.

The choice of model and training regime also ought to be clearly justified in the TEA case. Too often models and training regimes are chosen because they were used by researchers investigating a similar problem and not for some other reason, e.g., because of their explainability or impact on fairness for different prediction-recipient groups. Indeed, the models for our AI-enabled CDSS, described in section 3.3, were chosen because they are among the most commonly used for Type 2 diabetes-related problems [9,33]. Testing and validation benchmarks were chosen in a similar manner for our AI-enabled CDSS. That is, the default metrics that accompany the Caret (classification and regression training) package in the programming language R were used to assess the models. In particular, the measures are Accuracy, i.e., "the percentage of correctly classified instances out of all instances" and Cohen's Kappa (or simply Kappa) which is a measure of accuracy normalized to a baseline of random chance on the dataset [33]. While there is *prima facie* nothing unfair about using popular models and default testing and validation metrics, their use stands in need of justification. And it is in the TEA case that this justification ought to be made clear.

Indeed, the entire development phase ought to be comprehensively documented and summarized in the model documentation stage of the lifecycle. Here too fairness considerations arise. The *appropriateness* of the mode or format of documentation to affected stakeholders' needs and epistemic competencies, for example, has consequences for their understanding of the system. There are important links here between fairness and explainability for instance.

Consider that it may be difficult to justify the fairness of an AI-enabled CDSS if a clinician does not understand how or why the CDSS arrived at the prediction that it did. Evidence that information about the model is presented in concise and easily accessible terminology can both help justify its fairness and inspire trust in the system. To make our system more explainable, the importance of each of the 20 features used to train the models was calculated to show each feature's weighed importance levels when predicting the output [39]. Ideally, even decisions about chosen explainability methods should be explicitly stated and justified in the TEA case because different explainability methods, e.g., feature relevance explanations or explanation by example approaches, can mask unfair predictions an AI-enabled system might make [22].

*4.2.3 Fairness Assurance: The Deployment Phase*

While most research on AI-enabled systems ends in the development phase, such systems are increasingly developed for commercial reasons and are therefore deployed into existing activities and practices to operate alongside other technological systems and humans. That is the intention for the AI-enabled CDSS, that it will ultimately be deployed and used in clinical settings to aid clinicians in making (more) accurate diagnoses of hypertension in patients with Type 2 diabetes. And as with the previous two phases, there are important fairness claims that require justification in the deployment phase. For example, without input from clinicians in the design phase, the implementation of an AI-enabled CDSS might clash with existing sociotechnical practices with which clinicians, healthcare providers and patients are familiar. The burden may therefore fall, unfairly, on clinicians or other affected stakeholders to incorporate the AI-enabled CDSS into their existing workflow and at best this could result in unuse or misuse of the CDSS.

A similar concern arises in the user training stage, to wit, poor user training or a lack thereof may place an unfair burden on clinicians, as the expert users, to teach themselves how to operate and interact with the CDSS. Clinicians poorly trained to use the CDSS may not only mistrust the system, but that mistrust could have ramifications on patient health outcomes. Moreover, as highlighted in the fairness considerations map, it is at this stage that the respective roles of the clinician and AI-enabled CDSS ought to be clearly and unambiguously outlined. For example, is the CDSS's prediction akin to a second opinion from a junior colleague? A senior colleague? Will the clinician be punished if they disagree with the CDSS's prediction? What if the clinician disagrees with the CDSS and they are right? What if the clinician disagrees with the CDSS and they are wrong [46]? Failure to address questions such as these can result in clinicians shouldering an unfair burden of responsibility. Indeed, there are serious concerns that the human-in-the-loop working alongside an AI-enabled system in any context, be it a clinician or safety driver in an autonomous vehicle, will serve as a moral crumple zone or liability sink that absorbs moral responsibility and/or liability for whole system malfunctions when things go wrong [27].

Fairness is also of concern in the final two stages of the deployment phase. As alluded to above, an AI-enabled CDSS will inevitably change how clinicians interact with patients and how healthcare providers manage their services and resources. It will therefore be necessary to collect evidence that the CDSS continues to meet certain fairness goals or standards *while it is in use*. This is important for offline AI-enabled systems, i.e., systems which do not change throughout their use, but especially important for online AI-enabled systems, i.e., systems which change throughout their use, usually because some kind of ML is being employed to optimize its performance. Failure to properly monitor the CDSS in use can result in discriminatory outcomes for different patient groups. Similarly, failure to specify whose responsibility it is to monitor the system can place unfair burdens on clinicians since they may be expected to perform this task given their close proximity to the CDSS. Importantly, while there is in principle nothing unfair about having clinicians monitor the performance of the CDSS, why it is fair for them to take on such a role would have to be extensively justified in the TEA case. Perhaps clinicians will be compensated in some way for taking on this task, or will be discharged from carrying out certain other duties so that they have the time to monitor the CDSS. Whatever the justification might be, it will have to appear in the TEA case alongside evidence that clinicians can reliably fill this system monitoring role.

The concerns just raised for the system use and monitoring stage also largely apply to the final stage, system updating and deprovisioning. In short, justifying claims that the CDSS is fair requires evidence to demonstrate that an appropriate and qualified individual or team can initiate the update process. Additionally, evidence would be required to demonstrate that appropriate update and deprovisioning thresholds have been set. As in the previous stage, expecting clinicians to alert the relevant party that the system is in need of updating or deprovisioning may not only place an unfair burden on clinicians but it can impact how they treat their patients. For example, clinicians might be distracted by the performance of the CDSS and divert attention away from diagnosing their patient and instead focus on analyzing the performance of the CDSS. Furthermore, clinicians may lack access to the overall performance of the system and so be unable to say whether it is in need of updating because of a drift in performance, for example.

## 5 CONCLUSION

Though much attention has been devoted to the topic of AI fairness, most current approaches suffer from two limitations. They are narrow in scope both in terms of their substantive content and in terms of specifying when in the AI lifecycle issues of fairness ought to be considered or addressed. We have argued that one way these limitations can be addressed is through the use of TEA in order to systematically identify, document, justify and address considerations of fairness across the AI lifecycle. To illustrate the application of TEA, we have examined in detail the use of an AI-enabled CDSS that aids clinicians in diagnosing hypertension in Type 2 diabetes patients. Furthermore, to help facilitate the process of assuring fairness across the AI lifecycle, we have developed an open-source tool and a fairness considerations map to help affected stakeholders both develop their own assurance arguments and reflect upon salient fairness considerations in the healthcare context respectively. Ultimately, we hope that these contributions represent one step towards the cultivation of a flourishing assurance ecosystem in which richer socio-political considerations of fairness across the AI lifecycle are identified, documented and managed.